\documentstyle[prl,multicol,aps]{revtex}
\begin{document}
\input{epsf.tex}
\epsfverbosetrue

\title{Linear and nonlinear waveguides induced by optical vortex solitons}

\author{Andreas H. Carlsson, Johan N. Malmberg, Dan Anderson, and Mietek
Lisak}
\address{Department of Electrodynamics, Chalmers University of
Technology, S-412 96 G{\"o}teborg, Sweden}
\author{Elena A. Ostrovskaya, Tristram J. Alexander, and Yuri S. Kivshar}
\address{Optical Sciences Centre, Australian National University, Canberra ACT 0200, Australia}

\maketitle

\begin{abstract}
We study, numerically and analytically, linear and nonlinear waveguides induced by optical vortex solitons in a Kerr medium. Both fundamental and first-order guided modes are analyzed, as well as the cases of effectively defocusing and focusing nonlinearity.
\end{abstract}


\begin{multicols}{2}
\vspace{-2.0cm}
\narrowtext
Optical vortex solitons are stationary nonlinear beams self-trapped in two spatial dimensions and carrying a phase singularity - a nonlinear analog of wave-front dislocations \cite{review}.
 One of the possible applications of vortex solitons, as well as other types of spatial optical solitons, is to induce stable non-diffractive steerable waveguides that can guide and steer another beam, thus creating a reconfigurable all-optical circuit. 

In spite of many theoretical (see, e.g., \cite{shih,elena}) and experimental \cite{exp}
results on the soliton-induced waveguides in planar structures, waveguiding properties of optical solitons in a bulk medium have
not  yet been studied in detail. It is important to note that dark solitons are more attractive for waveguiding applications due to their greater stability and steerability \cite{review}. Their two-dimensional generalization, optical vortices, may have a number of potential applications in trapping small particles \cite{grover} as well as waveguiding. Preliminary numerical results \cite{sh_h_ol,law} indicate that some of the vortex waveguiding properties should be similar to those of planar dark solitons, however new features are expected. 

In this Letter we carry out a systematic analysis of the waveguides created by vortex solitons in a Kerr medium. We examine all the regimes. In the linear regime, the guided beam is weak and the analysis of waveguiding properties is possible by approximate analytical methods (see also Ref. \cite{law}). The nonlinear regime corresponds to the larger intensities of the guided beam and it gives rise to composite (or vector) solitons with a vortex component, identified and analyzed numerically. The effect of a focusing nonlinearity on a guided mode is also investigated.

We consider the stationary evolution of two incoherently coupled beams with
the frequencies $\omega_1$ and $\omega_2$ propagating in a bulk Kerr medium. Evolution of the slowly varying beam envelopes $E_1$
and $E_2$ is described by the normalized equations \cite{new}
\begin{eqnarray}\label{nls}
i \frac{\partial E_1}{\partial z} + \Delta_{\perp} E_1 -
(\eta^{-1}|E_1|^2 + \sigma|E_2|^2)E_1 = 0, \\ \nonumber
i \kappa \frac{\partial E_2}{\partial z} + \Delta_{\perp}
E_2 - \gamma (\eta |E_2|^2 + \sigma|E_1|^2)E_2 = 0, 
\end{eqnarray}
where $\Delta_{\perp}$ is the transverse Laplacian, $\eta = \omega_2^2/\omega_1^2$, $\kappa =k_2/k_1$, and
$k_{1,2}=2\pi/\lambda_{1,2}$ is the propagation constant of the beam with the frequency $\omega_{1,2}$. The field amplitudes and the propagation
distance $z$
are measured in the units of $k_1(n_0|n_2|)^{1/2}$ and $k_1n_0$,
respectively, $n_0$ is the linear refractive
index, and $n_2$ characterizes an intensity-dependent change of the refractive index, $\Delta n= n_2 |E|^2$.

Equations (\ref{nls}) can describe two different physical situations:
 the interaction of beams of two different frequencies ($\kappa$ and $\eta$ are
arbitrary), and the interaction of two orthogonally, linearly, or circularly polarized beams of the same
frequency ($\eta=\kappa=1$). In this Letter we consider the latter case. We assume that the beam $E_1$ exhibits
defocusing nonlinearity, whereas the beam $E_2$ exhibits either defocusing
$(\gamma = +1)$ or focusing $(\gamma = -1)$ nonlinearity, the latter may
occur for two orthogonally polarized beams in a photorefractive medium
\cite{ol}.

The parameter $\sigma>0$ measures the relative strength of
cross- and self-phase modulation effects. In general, its value
depends on the physical nature of nonlinearity, anisotropy of the
medium, and polarization state of the beams. For example, for linearly polarized beams in a Kerr-type material with a 
non-resonant electronic nonlinearity and cubic anisotropy, $\sigma = 2(2 \chi_{xxyy} + \chi_{xyyx})/3 \chi_{xxxx}$, $\chi_{ijkl}$ being
the elements of the third-order susceptibility tensor. In isotropic
materials $\chi_{xxxx} = 2 \chi_{xxyy} + \chi_{xyyx}$ and the values of $\sigma$ are fixed, namely $\sigma = 2/3$ and $\sigma = 2$ for linearly and circularly polarized beams, respectively. Other values of $\sigma$ are possible, for instance in the case of nonlinearity due to molecular orientation $\sigma=7$ \cite{boyd}.

We assume that the envelope $E_1$ carries a single-charged vortex on a
nonvanishing background field $|E_1| = E_0$, whereas the field $E_2$ is spatially localized, and it describes
either the fundamental or first-order mode guided by the vortex. Then, we look for
radially symmetric solutions in the form:
\begin{eqnarray}
\label{E}
E_1(R,\varphi;z)= \eta^{1/2} E_0 u(R) e^{-i{E_0}^2z} e^{in \varphi}, \\ \nonumber
E_2(R,\varphi;z)= \eta^{-1/2} E_0 v(R) e^{-i(\lambda/\kappa) {E_0}^2z}
 e^{im \varphi},\nonumber
\end{eqnarray}
where $R=\sqrt{x^2+y^2}$ and $n,m$ are topological charges. Coupled equations for the real $u$ and $v$ become
\begin{eqnarray}
\label{n nls}
 \Delta_{\rm r} u - \frac{n^2}{r^2} u + u - [u^2 +
(\sigma/\eta) v^2] u = 0, \\ \nonumber
\Delta_{\rm r} v - \frac{m^2}{r^2} v + \lambda v - \gamma
[v^2 + \eta \sigma u^2]v = 0, 
\end{eqnarray}
where $\Delta_{\rm r} \equiv d^2/dr^2 + (1/r)d/dr$ is the radial part of the
Laplacian, and $r = R E_0$. For any $\sigma$ and $\lambda$, Eqs. (\ref{n nls})  admit a solution for the vortex soliton in the form: $u=u_0(r)$, $v=0$. Here we are mainly interested in the case of a single charged vortex, thus $n=\pm 1$. For a fixed $\sigma$ and $\eta$, the family of composite
solitons $(u,v)$ formed by a vortex and its guided mode is described by
a single parameter $\lambda$. For $|v|/|u|=\varepsilon \ll 1$, the second equation of (\ref{n nls}) becomes a linear
eigenvalue problem with an effective potential created by the vortex. For $\gamma=+1$, the potential is {\em attractive}, and it may
support spatially localized solutions - {\em guided modes}, each of them
appearing at a certain cutoff value of $\lambda$. For $\gamma=-1$, the
potential is {\em repulsive} (``anti-waveguiding case''), and no guiding is
possible in the linear limit.
\begin{figure}
\setlength{\epsfxsize}{8cm}
\centerline{
\epsfbox{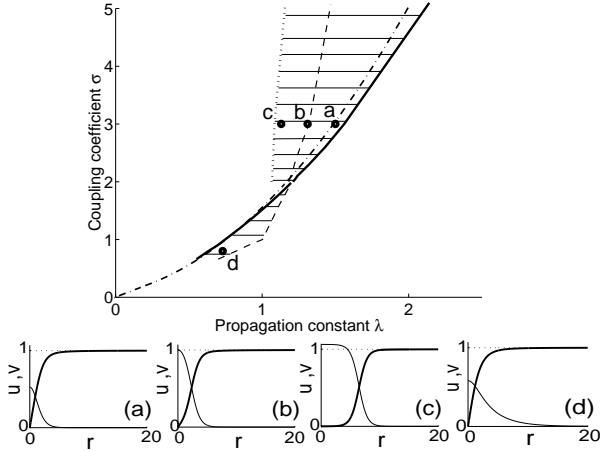}}
\caption{Existence domain (shaded) for the fundamental mode guided by a vortex. Points a, b, c, and d correspond to the modal structure examples (a-d) shown below. Parameters are: (a) - (c) $\sigma = 3$, and (d) $\sigma =0.8$. Dash-dotted - the cut-off line calculated by variational approach. }
\label{fig1}
\end{figure}
Away from the cut-off curve, the amplitude of the guided mode grows, and the linear waveguide approximation is no longer valid. In this nonlinear
waveguiding regime \cite{elena}, the guided mode deforms the supporting vortex waveguide, and together they form a composite, vector soliton. To identify the families of such solitons, we find, numerically, localized solutions of (\ref{n nls}) with the asymptotics: $u \rightarrow 1$, $v \rightarrow 0$ as $r \rightarrow \infty$. As has been previously established for $\sigma=2$\cite{sh_h_ol}, there exist different solutions of this type, classified by the order of the guided mode. Here we describe the entire existence domain for such solutions in the physically relevant region of parameters.

Several examples of the coupled-mode solutions for $\gamma=+1$ and $m=0$,
i.e. when the vortex guides a fundamental mode, are shown in Figs. 1(a-d). The regions of existence of all possible solutions of this type are also presented
in Fig. 1 on the parameter plane $(\sigma,\lambda)$, where the points (a) to (d) correspond to the profiles below.

In the parameter plane $(\lambda, \sigma)$, the cutoff of the fundamental mode is characterised by a curve $\sigma(\lambda)$ (thick curve, Fig. 1). Near this  curve, the amplitude of the guided mode is small and the linear waveguiding regime is valid. Linearizing Eqs. (\ref{n nls}) around the vortex solution, we obtain two decoupled equations:
\begin{eqnarray}
\label{eigen}
\Delta_r u_0 - \frac{n^2}{r^2} u_0 - u_0 + u^3_0 = 0,\\ \nonumber
\Delta_r v - \frac{m^2}{r^2} v - \lambda v + W_0(r) v = 0,
\end{eqnarray}
where only the terms $\sim \varepsilon$ are retained in the equation for the guided mode, and the effective guiding potential $W_0(r) = -\sigma u^2_0(r)$ is defined by the vortex shape. Since the vortex profile $u_0(r)$ is
not known in an explicit form, exact analytical solution of the eigenvalue problem for the mode $v(r)$ is not available. However, an estimate for the guided-mode cutoff can be obtained by a standard variational technique. First, we determine the vortex shape using a simple trial function $u_0(r) ={\rm tanh}^n(ar)$ and an effective Lagrangian for the first equation of (\ref{eigen}):
\begin{equation}
L_u=r \left( \frac{d u_0}{d r}\right)^2 + \frac{n^2}{r^2} u^2_0 + \frac{r}{2}(1+u^2_0)^2.
\end{equation}
The averaged Lagrangian is then found as the integral $\langle L \rangle = a^{-1}\int^{aR}_0 L_u(x) dx$ where $x=ar$, calculated in the limit $R\to \infty$. Variation of $\langle L_u \rangle$ with respect to $a$ yields the result: $a=n^{-1} \left\{\int^\infty_0 x [1-{\rm tanh}^{2n}(x)]^2 dx\right\}$, valid for any  $n$. This variational result provides the best fit to the numerical solution for $n\leq 3$. Then, the shape of the effective potential created by a single-charge vortex is determined by $u^2_0(r)={\rm tanh}^2(0.554 r)\approx 1-\exp(-r^2/4)$.

In order to find the cutoff for the fundamental mode guided by the vortex from the second equation of (\ref{eigen}), we take into account the $\varepsilon^2$-order correction to the vortex-induced waveguide due to a nonzero $v$-amplitude $W(r) = W_0+v^2(r)$. Assuming a trial function of the form $v(r)=B \exp(-r^2/b^2)$, and varying the averaged effective Lagrangian for the $v$-mode,
\begin{equation}
L_v=-r\left(\frac{dv}{dr}\right)^2+\left(\lambda r - \sigma u^2_0 r -\frac{r v^2}{2}\right) v^2,
\end{equation} 
with respect to $b$ and $B$, we obtain a simple analytical expression for the cut-off:
\begin{equation}
\sigma=\frac{1}{32}\left[1+24 \lambda + 16 \lambda^2 - (1-4 \lambda)\sqrt{1-4\lambda)^2}\right].
\end{equation}
A reasonably good agreement of this result with the numerically calculated cut-off can be seen in Fig. 1.

The properties of the vortex-induced waveguides and the corresponding composite solitons are different depending on whether $\sigma < 2$ or $\sigma > 2$ [see Fig. 1]. In the latter case, the amplitude of the guided mode grows
 with decreasing $\lambda$ [see Figs. 1(a,b)] and the vortex waveguide becomes broader, as in the case
of one-dimensional dark solitons \cite{elena}. The amplitude of the guided
mode reaches the vortex background far from the cutoff (dashed curve, Fig. 1) and
goes above the background value before the solution expands and disappears
(dotted curve, Fig. 1). Close to this limit the solution corresponds to a type of polarization domain walls
\cite{sh_h_ol}. Its expansion with changing $\lambda$ is very rapid, and hence the border of the existence region (dotted curve, Fig. 1) is determined by the accuracy of numerical integration. For $\sigma < 2$ the scenario is similar as $\lambda$ increases but the amplitude never reaches the background before the
solution disappears on the other border (dashed curve, Fig. 1).
\begin{figure}
\setlength{\epsfxsize}{6.5cm}
\centerline{
\epsfbox{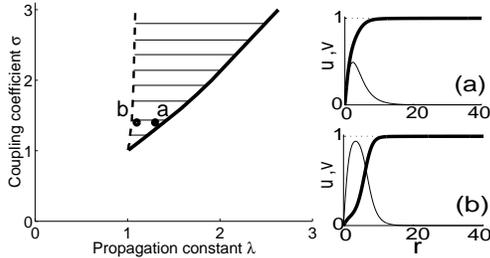}}
\caption{Existence domain (shaded) and characteristic examples
for (a) a first-order mode guided by a vortex and (b) corresponding vector soliton.}
\label{fig2}
\end{figure}

\begin{figure}
\setlength{\epsfxsize}{6cm}
\centerline{
\epsfbox{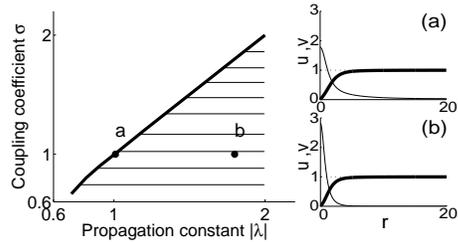}}
\caption{Existence domain of vector solitons (shaded) and two examples (a,b) of the vector solitons in the anti-waveguiding regime.}
\label{fig3}
\end{figure}

Next, we examine the guiding of the first-order mode by an optical vortex, i.e.
we take  $m = \pm 1$ in  Eqs. (\ref{n nls}). The existence domain for such
solutions is shown in Fig. 2 (left). For $\sigma = 1$, Eqs.
(\ref{n nls}) represent a two-dimensional generalization of a completely
integrable one-dimensional Manakov system \cite{manakov}. In this case,
waveguiding properties of a vortex resemble
those of a one-dimensional Manakov dark soliton. Namely, the induced
waveguide is single-mode, and the first-order mode exists only at $\lambda
= \sigma = 1$.

Although for $\sigma > 1$ the vortex-induced waveguide is expected to guide
higher-order modes, the corresponding composite solitons are likely to be
dynamically unstable at least in the nonlinear regime shown in Fig. 2(b).
 As $\lambda$ decreases away from the cutoff, the amplitude of the guided
mode grows and the waveguide becomes broader [see Fig. 2(b)] until its
shape resembles that of a radially symmetric domain wall \cite{sh_h_ol}. The
solution disappears when its localization region is no longer finite.

At last, we briefly discuss the case of a focusing nonlinearity, i.e. $\gamma=-1$ in Eqs. (\ref{nls}). Although no bound modes of the vortex anti-waveguide can exist in the linear regime $(|v| \ll |u|)$, the numerical results suggest that for large intensities of the bright component mutual trapping is still possible. This effect is similar to nonlinear anti-waveguides assisted by an external potential \cite{gisin}.
 Stationary solutions shown in Fig. 3 (right) for a fixed
$\sigma$, appear when an absolute value of $\lambda$ ($\lambda < 0$
for $\gamma =-1$) exceeds a certain threshold value (solid curve, Fig. 3). Away from this border, the amplitude of the bright soliton increases. 

Our results can be readily generalized to the case of beams of different colours, $\eta \neq 1$. For the limit of linear waveguiding, the effective coupling constant changes to $\eta \sigma$ [see Eqs. (\ref{n nls})], and the properties of the vortex waveguide become dependent on the frequency ratio.

In conclusion, we have analysed the waveguiding properties of optical vortex solitons in Kerr media and shown that they crucially depend on the relative strength of the cross- and self-phase modulation effects. We have identified families of composite solitons formed by a vortex and its guided mode with or without a topological charge and introduced a simple technique of determining the cutoff of the guided mode by a variational approach which can be applied to other problems of nonlinear waveguiding.

\vspace{-2mm}
\end{multicols}

\end{document}